\documentclass{appolb}
\usepackage{graphicx}
\begin{document}
\newcommand{\PrOsSb}{PrOs$_4$Sb$_{12}$}
\newcommand{\LaOsSb}{LaOs$_4$Sb$_{12}$}
\hyphenation {Schott-ky} \hyphenation {su-per-con-duc-tor}
\hyphenation {su-per-con-duc-ting} \hyphenation
{su-per-con-duc-tiv-i-ty} \hyphenation {quad-ru-pole} \hyphenation
{quad-ru-po-lar} \hyphenation {skut-ter-u-dite} \hyphenation
{an-ti-fer-ro-mag-net-ic} \hyphenation {pre-fac-tor} \hyphenation
{mul-ti-plet}

\title{A new heavy fermion superconductor: the filled skutterudite
compound \PrOsSb
\thanks{Presented at the Strongly Correlated Electron Systems
Conference, Krak\'ow 2002}%
}


\author{M. B. Maple, P.-C. Ho, N. A. Frederick, V. S. Zapf, W. M. Yuhasz,
and E. D. Bauer
\address{Department of Physics and Institute for Pure and
Applied Physical Sciences, University of California, San Diego,
La Jolla, CA 92093 USA}
}
\maketitle


\begin{abstract}

The filled skutterudite compound \PrOsSb{} exhibits
superconductivity below a critical temperature $T_\mathrm{c} =
1.85$ K that develops out of a nonmagnetic heavy Fermi liquid with
an effective mass $m^{*} \approx 50~m_\mathrm{e}$, where
$m_\mathrm{e}$ is the free electron mass. Analysis of magnetic
susceptibility, specific heat, electrical resistivity and
inelastic neutron scattering measurements within the context of a
cubic crystalline electric field yields a Pr$^{3+}$ energy level
scheme that consists of a $\Gamma_{3}$ nonmagnetic doublet ground
state that carries an electric quadrupole moment, a low lying
$\Gamma_{5}$ triplet excited state at $\sim 10$ K, and
$\Gamma_{4}$ triplet and  $\Gamma_{1}$ singlet excited states at
much higher temperatures.  The superconducting state appears to be
unconventional and to consist of two distinct superconducting
phases.  An ordered phase of magnetic or quadrupolar origin occurs
at high fields and low temperatures, suggesting that the
superconductivity may occur in the vicinity of a magnetic or
electric quadrupolar quantum critical point.

\end{abstract}



\section{Introduction}

The filled skutterudite compound \PrOsSb{} has been reported to
exhibit superconductivity below a critical temperature
$T_\mathrm{c} = 1.85$ K that develops out of a heavy Fermi liquid
with an effective mass $m^{*} \approx 50~m_\mathrm{e}$, where
$m_\mathrm{e}$ is the free electron mass
\cite{Maple01,Bauer02a,Maple02,Ho01} .  To our knowledge,
\PrOsSb{} is the first heavy fermion superconductor based on Pr;
all of the other known heavy fermion superconductors (about $20$)
are compounds of Ce or U.  The superconducting state appears to be
unconventional in nature and may consist of two distinct
superconducting phases  \cite{Maple02,Izawa02}.   An ordered
phase, presumably of magnetic or quadrupolar origin, occurs at
high fields $> 4.5$ T and low temperatures $< 1.5$ K
\cite{Maple02,Ho01,Vollmer02,Oeschler02,Tenya02,Aoki02},
suggesting that the superconductivity may occur in the vicinity of
a magnetic or quadrupolar quantum critical point (QCP).   Analysis
of magnetic susceptibility $\chi{}(T)$, specific heat $C(T)$,
electrical resistivity $\rho(T)$ and inelastic neutron scattering
measurements within the context of a cubic crystalline electric
field (CEF) yields a Pr$^{3+}$ energy level scheme that consists
of a $\Gamma_{3}$ nonmagnetic doublet ground state that carries an
electric quadrupole moment, a low lying $\Gamma_{5}$ triplet
excited state at $\sim 10$ K, and $\Gamma_{4}$ triplet and
$\Gamma_{1}$ singlet excited states at much higher temperatures
($\sim 130$ K and $\sim 313$ K, respectively)
\cite{Maple01,Bauer02a,Maple02,Ho01}.  This scenario suggests that
the underlying mechanism of the heavy fermion behavior in
\PrOsSb{} may involve the interaction of  Pr$^{3+}$ electric
quadrupole moments with the charges of the conduction electrons,
rather than the interaction of  Pr$^{3+}$ magnetic dipole moments
with the spins of the conduction electrons, the interaction that
is widely believed to be responsible for the heavy fermion state
in most Ce- and U-based compounds.  It also raises the possibility
that electric quadrupole fluctuations play a role in the
superconductivity of \PrOsSb.  In this paper, we briefly review
the current experimental situation regarding the heavy fermion
state, the superconducting state, and a high field, low
temperature phase that is apparently associated with magnetic or
electric quadrupolar order in  \PrOsSb.

\section{Heavy fermion state in \PrOsSb}

\begin{figure}[!ht]
\begin{center}
\includegraphics[width=0.6\textwidth]{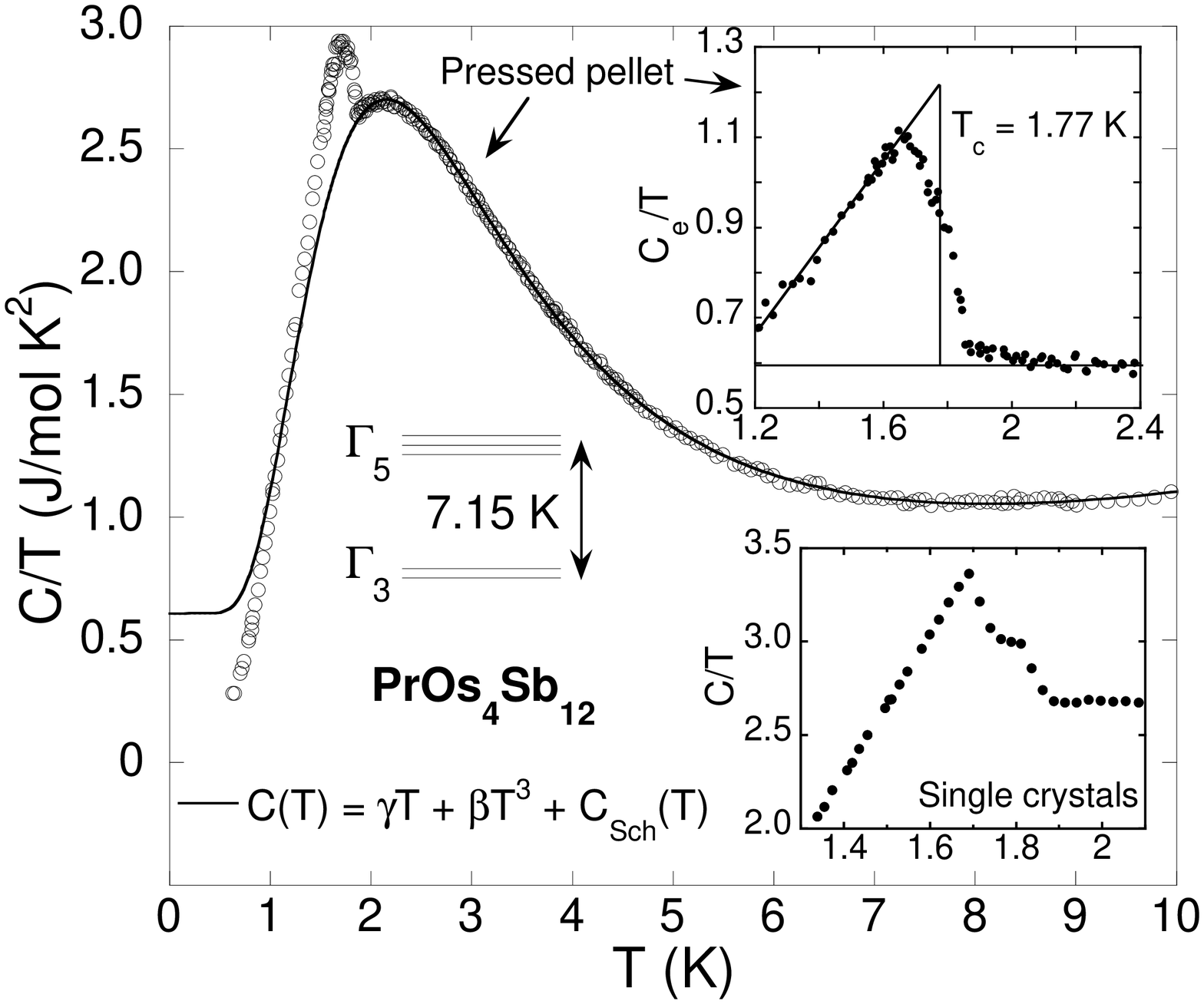} 
\end{center}
\caption{Specific heat $C$ divided by temperature $T$, $C/T$, vs
$T$ for a \PrOsSb{} pressed pellet.  The line represents a fit of
the sum of electronic, lattice, and Schottky contributions to the
data.  Upper inset: $C_\mathrm{e}/T$ vs $T$ near $T_\mathrm{c}$
for a \PrOsSb{} pressed pellet ($C_\mathrm{e}$ is the electronic
contribution to $C$).  Lower inset: $C/T$ vs $T$ near
$T_\mathrm{c}$ for \PrOsSb{} single crystals, showing the
structure in $\Delta C$ near $T_\mathrm{c}$. Data from
Ref.~\cite{Maple01,Bauer02a}.}
\label{heat}
\end{figure}

The first evidence for a heavy fermion state in the filled
skutterudite compound \PrOsSb{} emerged from specific heat $C(T)$
measurements on a \PrOsSb{} pressed pellet (formed by pressing a
collection of small single crystals in a cylindrical die) at low
temperatures.  Specific heat data  in the form of a plot of $C/T$
vs $T$ between $0.5$ K and $10$ K  for the \PrOsSb{} pressed
pellet from Refs. \cite{Maple01} and \cite{Bauer02a} are shown in
Fig.~\ref{heat}.  The $C(T)$ data have been corrected for excess
Sb derived from the molten Sb flux in which the crystals were
grown.  The line in the figure represents the expression $C(T) =
\gamma T + \beta T^{3} + C_\mathrm{Sch}(T)$, where $\gamma T$ and
$\beta T^{3}$ are electronic and phonon contributions,
respectively, and $C_\mathrm{Sch}(T)$ is a Schottky anomaly for a
two level system consisting of a doublet ground state and a
triplet excited state at an energy $\Delta$ above the ground
state.  The best fit of this expression to the data yields the
values $\gamma$ = $607$ mJ/mol K$^{2}$, $\beta = 3.95$ mJ/mol
K$^{4}$ (corresponding to a Debye temperature $\theta_\mathrm{D} =
203$ K), and $\Delta = 7.15$ K. Superimposed on the Schottky
anomaly is a feature in the specific heat due to the onset of
superconductivity at $T_\mathrm{c} = 1.85$ K which is also
observed as an abrupt drop in $\rho(T)$ to zero and as a sharp
onset of diamagnetism in $\chi(T)$. The feature in $C(T)/T$ due to
the superconductivity is also shown in the top inset of
Fig.~\ref{heat} along with an entropy conserving construction from
which the ratio of the jump in specific heat $\Delta C$ at
$T_\mathrm{c}$, $\Delta C/T_\mathrm{c} = 632$ mJ/mol K$^{2}$, has
been estimated. Using the BCS relation $\Delta
C/\gamma{}T_\mathrm{c} = 1.43$, this yields a value for $\gamma$
of $440$ mJ/mol K$^{2}$.  The value of $\Delta C/T_\mathrm{c}$ is
larger than that reported in Ref.~\cite{Bauer02a} due to the
correction of the $C(T)$ data for the excess Sb (about $30$
percent of the total mass). This value is comparable to that
inferred from the fit to the $C/T$ vs $T$ data in the normal state
above $T_\mathrm{c}$, and indicative of heavy fermion behavior.  A
similar analysis of the $C(T)$ data taken at the University of
Karlsruhe on several single crystals of \PrOsSb{} prepared in our
laboratory yielded $\gamma = 313$ mJ/mol K$^{2}$,
$\theta_\mathrm{D} = 165$ K, $\Delta = 7$ K, and $\Delta
C/\gamma{}T_\mathrm{c} \approx 3$, much higher than the BCS value
of $1.43$ and indicative of strong coupling effects
\cite{Vollmer02}. It is interesting that we also find a large
value of $\Delta C/\gamma{}T_\mathrm{c} \approx 3$ in recent
$C(T)$ measurements on one single crystal of \PrOsSb{} at UCSD.
Although the values of $\gamma$ determined from these experiments
vary somewhat, they are all indicative of a heavy electron ground
state and an effective mass $m^{*} \approx 50~m_\mathrm{e}$.

\begin{figure}[!ht]
\begin{center}
\includegraphics[width=0.6\textwidth]{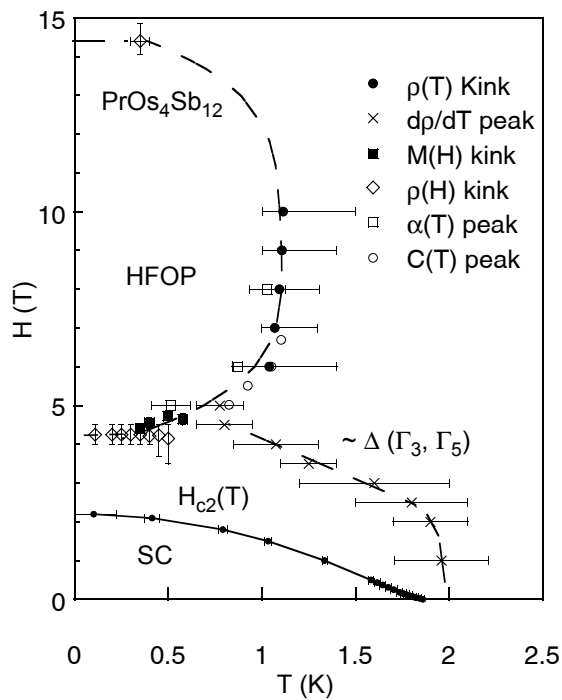} 
\end{center}
\caption{Magnetic field - temperature ($H - T$) phase diagram of
\PrOsSb{} showing the regions exhibiting superconductivity (SC)
and the high field ordered phase (HFOP). The dashed line is a
measure of the splitting between the Pr$^{3+}~\Gamma_{3}$ ground
state and $\Gamma_{5}$ excited state (see text for further
details). Data from
Refs.~\cite{Maple02,Ho01,Vollmer02,Oeschler02,Ho02}.}
\label{phase}
\end{figure}

Further evidence of heavy fermion superconductivity is provided by
the upper critical field $H_\mathrm{c2}$ vs $T$ curve which is
shown in Fig.~\ref{phase} \cite{Bauer02a,Maple02}.  The  orbital
critical field $H^{*}_\mathrm{c2}(0)$ can be derived from the
slope of the $H_\mathrm{c2}$ curve near $T_\mathrm{c}$ and used to
estimate the superconducting coherence length $\xi_{0} \approx$
\mbox{$116$ \AA{}} via the relation $H^{*}_\mathrm{c2}(0) =
\Phi_{0}/2\pi \xi_{0}^{2}$, where $\Phi_{0}$ is the flux quantum.
The Fermi velocity $v_\mathrm{F}$ can be obtained from the BCS
relation $\xi_{0} = 0.18 \hbar
v_\mathrm{F}/k_\mathrm{B}T_\mathrm{c}$ and used to determine the
effective mass $m^{*}$ by means of the expression $m^{*} = \hbar
k_\mathrm{F}/v_\mathrm{F}$.  Using a simple free electron model to
estimate the Fermi wave vector $k_\mathrm{F}$, an effective mass
$m^{*} \approx 50~m_\mathrm{e}$ is obtained
\cite{Bauer02a,Maple02}. Calculating $\gamma$ from $m^{*}$ yields
$\gamma \sim 350$ mJ/mol K$^{2}$, providing further evidence for a
heavy fermion state in \PrOsSb.

Recently, Sugawara et al.~\cite{Sugawara02} performed de Haas-van
Alphen effect measurements on \PrOsSb.  They found that the
topology of the Fermi surface is close to that of the reference
compound \LaOsSb{} and is explained well by band structure
calculations.  In contrast to the similarity in the Fermi surface
topologies of \PrOsSb{} and \LaOsSb{}, the cyclotron effective
masses $m_\mathrm{c}^{*}$ of \PrOsSb{} are up to $\sim 6$ times
enhanced compared to those of \LaOsSb. The Sommerfeld coefficient
$\gamma$ estimated from the Fermi surface volume and the value of
$m_\mathrm{c}^{*}$, assuming a spherical Fermi surface, is $\sim
150$ mJ/mol K$^{2}$, which is two to three times smaller than the
value of $\gamma$ inferred from the normal and superconducting
properties of \PrOsSb.  Our studies of \LaOsSb{} single crystals
reveal superconductivity with a $T_\mathrm{c}$ of $1$ K.

\section{The Pr$^{3+}$ energy level scheme in the crystalline
electric field}

\begin{figure}[!ht]
\begin{center}
\includegraphics[width=0.6\textwidth]{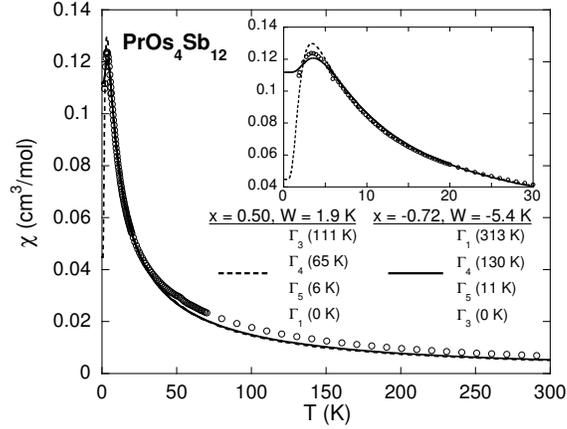} 
\end{center}
\caption{Magnetic susceptibility $\chi$ vs $T$ for \PrOsSb{}
single crystals.  Fits of a CEF model to the $\chi(T)$ data in
which the ground state is either a $\Gamma_{3}$ nonmagnetic
doublet (solid line) or a $\Gamma_{1}$ singlet (dashed line) are
indicated in the figure. Inset: $\chi(T)$ below $30$ K.  After
Ref.~\cite{Bauer02a}.} \label{chi}
\end{figure}

Magnetic susceptibility $\chi{}(T)$ data between $\sim 1$ K and
room temperature for  \PrOsSb{} from Ref.~\cite{Bauer02a} are
shown in Fig.~\ref{chi}.  These $\chi{}(T)$ data have been
corrected for excess Sb by assuming the effective moment
$\mu_\mathrm{eff}$ was equal to the full Hund's rule value of
$3.6~\mu_\mathrm{B}$.  This led to an estimation of the mass of
the excess Sb to be $\sim 25\%$ of the total mass. The $\chi{}(T)$
data exhibit a peak at $\sim 3$ K and saturate to a value of $\sim
0.11$ cm$^{3}$/mol as $T \rightarrow 0$, indicative of a
nonmagnetic ground state. At temperatures above $\sim 5$ K,
$\chi{}(T)$ is strongly T-dependent, as expected for well defined
Pr$^{3+}$ magnetic moments.  In the analysis of the $\chi{}(T)$
data, interactions between Pr$^{3+}$ ions and hybridization of the
Pr $4$f and conduction electron states were neglected, while the
degeneracy of the Hund's rule multiplet of the  Pr$^{3+}$ ions was
assumed to be lifted by a cubic crystalline electric field (CEF)
and to have a nonmagnetic ground state.  According to Lea, Leask,
and Wolf \cite{Lea62}, in a cubic CEF, the Pr$^{3+} J = 4$ Hund's
rule multiplet splits into a $\Gamma_{1}$ singlet, a $\Gamma_{3}$
nonmagnetic doublet that carries an electric quadrupole moment,
and $\Gamma_{4}$ and $\Gamma_{5}$ triplets.  It was assumed that
the nonmagnetic ground state of the Pr$^{3+}$ ions corresponds to
either a $\Gamma_{1}$ singlet or a $\Gamma_{3}$ nonmagnetic
doublet \cite{Bauer02a}. Although reasonable fits to the
$\chi{}(T)$ data could be obtained for both $\Gamma_{1}$ and
$\Gamma_{3}$ ground states, as shown in Fig.~\ref{chi}, the most
satisfactory fit was obtained for a $\Gamma_{3}$ nonmagnetic
doublet ground state with a $\Gamma_{5}$ first excited triplet
state at $11$ K and $\Gamma_{4}$ and $\Gamma_{1}$ excited states
at $130$ K and $313$ K, respectively (see the inset to
Fig.~\ref{chi}). Inelastic neutron scattering measurements on
\PrOsSb{} \cite{Maple02} reveal peaks in the INS spectrum at
$0.71$ meV ($8.2$ K) and $11.5$ meV ($133$ K) that appear to be
associated with transitions between the $\Gamma_{3}$ ground state
and the $\Gamma_{5}$ first and $\Gamma_{4}$ second excited states,
respectively, that are in good agreement with the Pr$^{3+}$ CEF
energy level scheme determined from the analysis of the
$\chi{}(T)$ data.  As noted above, the Schottky anomaly in the
$C(T)$ data on \PrOsSb{} taken at UCSD and at the University of
Karlsruhe \cite{Vollmer02} can be described well by a two level
system consisting of a doublet ground state and a low lying
triplet excited state with a splitting of $\sim 7$ K, a value that
is comparable to the values deduced from the $\chi{}(T)$ and INS
data. However, a $\Gamma_{1}$ ground state cannot, at this point,
be completely excluded.

\begin{figure}[!ht]
\begin{center}
\includegraphics[width=0.6\textwidth]{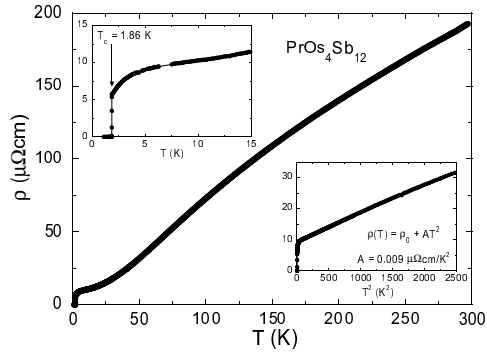} 
\end{center}
\caption{Electrical resistivity $\rho$ vs $T$ for \PrOsSb{}
between $1.8$ K and $300$ K.  Upper inset: $\rho(T)$ below $20$ K.
Lower inset: $\rho(T)$ below $50$ K.  After Ref.~\cite{Maple01}.}
\label{res}
\end{figure}

While a magnetic $\Gamma_{4}$ or $\Gamma_{5}$ Pr$^{3+}$ ground
state could also produce a nonmagnetic heavy fermion ground state
via an antiferromagnetic exchange interaction (Kondo effect), the
behavior of $\rho{}(T)$ of \PrOsSb{} in the normal state does not
resemble the behavior of $\rho{}(T)$ expected for this scenario.
For a typical magnetically-induced heavy fermion compound,
$\rho{}(T)$ often increases with decreasing temperature due to
Kondo scattering, reaches a maximum, and then decreases rapidly
with decreasing temperature as the highly correlated heavy fermion
state forms below the coherence temperature.  At low temperatures,
$\rho{}(T)$ typically varies as $AT^{2}$ with a prefactor $A
\approx 10^{-5}$ [$\mu{}\Omega$~cm~K$^{2}$(mJ/mol)$^{-2}$]
$\gamma^{2}$ that is consistent with the Kadowaki-Woods relation
\cite{Kadowaki86}.  In contrast,  $\rho{}(T)$ of \PrOsSb{}, shown
in Fig.~\ref{res} \cite{Maple01,Bauer02a}, exhibits typical
metallic behavior with negative curvature at higher temperatures
and a pronounced `roll off' below $\sim 8$ K before it vanishes
abruptly when the compound becomes superconducting (upper inset of
Fig.~\ref{res}). The temperature of the `roll off' in $\rho{}(T)$
is close to that of the decrease in charge or spin dependent
scattering due to the decrease in population of the proposed first
excited state ($\Gamma_{5}$) as the temperature is lowered.  The
$\rho{}(T)$ data are shown in the lower inset of Fig.~\ref{res}
and can be described by a temperature dependence of the form
$AT^{2}$ between $\sim 8$ K and $45$ K, but with a prefactor $A
\approx 0.009~\mu\Omega~$cm/K$^{2}$ that is nearly two orders of
magnitude smaller than that expected from the Kadowaki-Woods
relation ($A \approx 1.2~\mu\Omega$~cm/K$^{2}$ for $\gamma \approx
350$ mJ/mol K$^{2}$) \cite{Kadowaki86}. Interestingly, $\rho{}(T)$
is consistent with $T^{2}$ behavior with a value $A \approx
1~\mu\Omega$~cm/K$^{2}$ in fields of $\sim 5$ T \cite{Ho01} in the
high field ordered phase discussed in section $5$. The zero-field
temperature dependence of $\rho{}(T)$ is similar to that observed
for the compound PrInAg$_{2}$, which also has a low value of the
coefficient $A$, an enormous $\gamma$ of $\sim 6.5$ J/mol K$^{2}$,
and a $\Gamma_{3}$ nonmagnetic doublet ground state
\cite{Yatskar96}. The compounds \PrOsSb, PrInAg$_{2}$, and another
Pr-based skutterudite, PrFe$_{4}$P$_{12}$ \cite{Sato00}, may
belong to a new class of heavy fermion compounds in which the
heavy fermion state is produced by electric quadrupole
fluctuations.  In contrast, magnetic dipole fluctuations are
widely believed to be responsible for the heavy fermion state in
most Ce and U heavy fermion compounds (with the possible exception
of certain U compounds such as UBe$_{13}$). Another possible
source of the enhanced effective mass in \PrOsSb{} may involve
excitations from the ground state to the the low lying first
excited state in the Pr$^{3+}$ CEF energy level scheme
\cite{Fulde02}.

Two studies of the nonlinear magnetic susceptibility have been
performed in an attempt to determine the CEF ground state of the
Pr$^{3+}$ ion in \PrOsSb{} \cite{Tenya02,Bauer02b}. The nonlinear
susceptibility $\chi_{3}$ is the coefficient of an $H^{3}$ term in
the expansion of the magnetization $M$ in a series of odd powers
of $H$; i.e., $M \approx \chi_{1}H + (\chi_{3}/6)H^{3}$, where
$\chi_{1}$ is the ordinary linear susceptibility.  For an ionic
situation, $\chi_{3}$ is isotropic and varies as $T^{-3}$ for a
magnetic ground state, whereas for a non-Kramers $\Gamma_{3}$
doublet it is anisotropic and diverges at low temperatures for
$H~||~[100]$ and approaches a constant for $H~||~[111]$
\cite{Morin81}.  This type of study was previously employed in an
attempt to determine the ground state of U in the compound
UBe$_{13}$ \cite{Ramirez94}. In the study by Bauer et al.
\cite{Bauer02b}, $\chi_{3}$ was found to be anisotropic and
exhibit a minimum followed by a maximum and a negative divergence
as the temperature was decreased for $H~||~[100]$, while
$\chi_{3}$ exhibited a minimum and then increased down to the
lowest temperature of the measurement \mbox{($T = 1.8$ K)}, for
$H~||~[111]$. Comparison with calculations based on the
quadrupolar Anderson-Hamiltonian provided a qualitative
description of the $\chi_{3}(T)$ data for $H~||~[100]$, but did
not describe the $\chi_{3}(T)$ data well for $H~||~[111]$.  It was
concluded that the data were qualitatively consistent with a
$\Gamma_{3}$ ground state, given the limits of the experiment and
the complexity of the theory. On the other hand, a study by Tenya
et al. \cite{Tenya02} found $\chi_{3}(T)$ to be nearly isotropic.
It was concluded that the a $\Gamma_{1}$ ground state could not be
ruled out on the basis of this experiment. However, it should be
noted that these $\chi_{3}(T)$ studies are difficult to interpret
because of the curvature of $M(H)$ and the complications that
arise at lower temperatures $T \leq T_\mathrm{c}$ and lower fields
$H \leq H_\mathrm{c2}$ due to the superconductivity and at
temperatures $T \leq 2$ K and higher fields $H \geq 4.5$ K by the
onset of the high field ordered phase, discussed in section $5$.

\section{Superconducting state}

Several features in the superconducting properties of \PrOsSb{}
indicate that the superconductivity of this compound is
unconventional in nature.  First, $C_\mathrm{s}(T)$ follows a
power law T-dependence, $C_\mathrm{s}(T) \sim T^{2.5}$, after the
Schottky anomaly and $\beta T^{3}$ lattice contributions have been
subtracted from the $C(T)$ data. As reported in
Ref.~\cite{Maple02}, $C_\mathrm{s}(T)$ follows a power law with
$C_\mathrm{s}(T) \sim T^{3.9}$ when the Schottky anomaly is not
subtracted. Second, there is a `double-step' structure in the jump
in $C(T)$ near $T_\mathrm{c}$ in single crystals (lower inset of
Fig.~\ref{heat}) that suggests two distinct superconducting phases
with different $T_\mathrm{c}$'s: $T_\mathrm{c1} \approx 1.70$ K
and $T_\mathrm{c2} \approx 1.85$ K \cite{Maple02,Vollmer02}.  This
structure is not evident in the $C(T)$ data taken on the pressed
pellet of \PrOsSb{} shown in the upper inset of Fig.~\ref{heat},
possibly due to strains in the single crystals out of which the
pressed pellet is comprised that broaden the transitions at
$T_\mathrm{c1}$ and $T_\mathrm{c2}$ so that they overlap and
become indistinguishable. However, at this point, it is not clear
whether these two apparent jumps in $C(T)$ are associated with two
distinct superconducting phases or are due to sample
inhomogeneity.  It is noteworthy that all of the single crystal
specimens prepared in our laboratory and investigated by our group
and our collaborators exhibit this `double-step' structure.
Multiple superconducting transitions, apparently associated with
distinct superconducting phases, have previously been observed in
two other heavy fermion superconductors, UPt$_{3}$
\cite{Lohneysen94} and U$_{1-x}$Th$_{x}$Be$_{13}~(0.1 \leq x \leq
0.35)$ \cite{Ott85}. Measurements of the specific heat in magnetic
fields reveal that the two superconducting features shift downward
in temperature at nearly the same rate with increasing field,
consistent with the smooth temperature dependence of the
$H_\mathrm{c2}(T)$ curve \cite{Vollmer02}. These two transitions
have also been observed in thermal expansion measurements
\cite{Oeschler02}, which, from the Ehrenfest relation, reveal that
$T_\mathrm{c1}$ and $T_\mathrm{c2}$ have different pressure
dependencies, suggesting that they are associated with two
distinct superconducting phases.

Recent transverse field $\mu$SR \cite{MacLaughlin02} and Sb-NQR
measurements \cite{Kotegawa02} on \PrOsSb{} are consistent with an
isotropic energy gap. Along with the specific heat, these
measurements indicate strong coupling superconductivity. These
findings suggest an s-wave, or, perhaps, a Balian-Werthamer p-wave
order parameter. On the other hand, the superconducting gap
structure of \PrOsSb{} was investigated by means of thermal
conductivity measurements in magnetic fields rotated relative to
the crystallographic axes by Izawa et al. \cite{Izawa02}.  These
measurements reveal two regions in the $H-T$ plane, a low field
region in which $\Delta(\bf{k})$ has two point nodes, and a high
field region where $\Delta(\bf{k})$ has six point nodes.  The line
lying between the low and high field superconducting phases may be
associated with the transition at $T_\mathrm{c2}$, whereas the
line between the high field phase and the normal phase,
$H_\mathrm{c2}(T)$, converges with $T_\mathrm{c1}$ as $H
\rightarrow 0$.  Clearly, more research will be required to
further elucidate the nature of the superconducting state in
\PrOsSb.

\section{High field ordered phase}

Evidence for a high field ordered phase was first derived from
magnetoresistance measurements in the temperature range $80$ mK
$\leq T \leq 2$ K and magnetic fields up to $9$ tesla
\cite{Maple02,Ho02}. The $H-T$ phase diagram, depicting the
superconducting region and the high field ordered phase is shown
in Fig.  2.  The line that intersects the high field ordered state
represents the inflection point of the `roll-off' in $\rho(T)$ at
low temperatures and is a measure of the splitting between the
Pr$^{3+}$ ground state and the first excited state, which
decreases with field (see Fig.~\ref{res}).  The high field ordered
phase has also been observed by means of large peaks in the
specific heat \cite{Vollmer02,Aoki02} and thermal expansion
\cite{Oeschler02} and kinks in magnetization vs magnetic field
curves \cite{Tenya02,Ho02} in magnetic fields $> 4.5$ T and
temperatures $< 1.5$ K.

\begin{figure}[!ht]
\begin{center}
\includegraphics[width=0.6\textwidth]{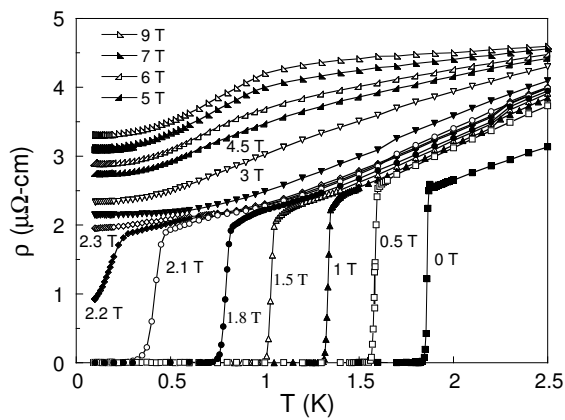} 
\end{center}
\caption{Electrical resistivity $\rho$ vs $T$ for various magnetic
fields up to $9$ T for a \PrOsSb{} single crystal.  The rapid drop
in $\rho(T)$ to zero for $H < 2.3$ T is due to the superconducting
transition, while the decrease in $\rho(T)$ for $H \geq 4.5$ T
below $\sim 1$ K appears to be due to a field induced phase,
presumably due to magnetic or quadrupolar order.  After
Ref.~\cite{Ho01}.} \label{magres}
\end{figure}

\begin{figure}[!ht]
\begin{center}
\includegraphics[width=0.6\textwidth]{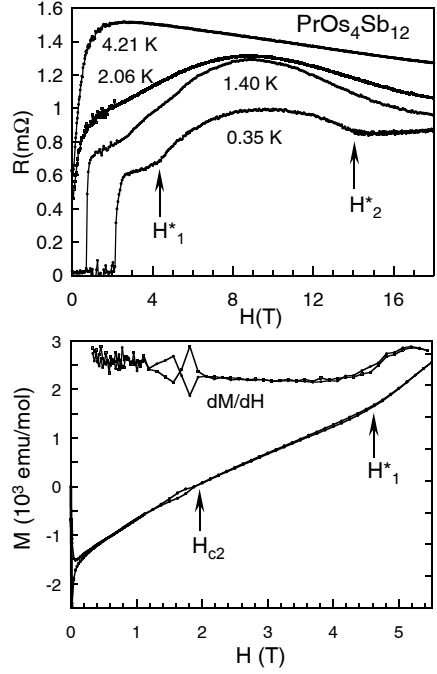} 
\end{center}
\caption{(a) Electrical resistance $R$ vs magnetic field $H$ for
\PrOsSb{} at several different temperatures below $4.21$ K for $0
\leq H \leq 18$ T. (b) Magnetization $M$ vs $H$ for \PrOsSb{} at
$0.34$ K for $0 \leq H \leq 5.5$ T.}
\label{HFOP}
\end{figure}

Shown in Fig.~\ref{magres} are $\rho{}(T)$ data for various
magnetic fields up to $9$ T for \PrOsSb, which reveal drops in
$\rho{}(T)$ due to the superconductivity for $H \leq 2.3$ T and
features in $\rho{}(T)$ associated with the onset of the high
field ordered phase for $H \geq 4.5$ T.  Isotherms of electrical
resistance $R$ vs $H$ and magnetization $M$ vs $H$ are shown in
Figs.~\ref{HFOP}(a) and \ref{HFOP}(b), respectively.  The fields
denoting the boundaries of the high field ordered phase,
$H_{1}^{*}$ and $H_{2}^{*}$, are indicated in the figure.

\section{Summary}

Experiments on the filled skutterudite compound \PrOsSb{} have
revealed a number of extraordinary phenomena: a heavy fermion
state characterized by an effective mass $m^{*} \approx
50~m_\mathrm{e}$, unconventional superconductivity below
$T_\mathrm{c} = 1.85$ K with two distinct superconducting phases,
and a high field ordered phase, presumably associated with
magnetic or electric quadrupolar order. Analysis of $\chi{}(T)$,
$C(T)$, $\rho{}(T)$, and INS data indicate that Pr$^{3+}$ has a
nonmagnetic $\Gamma_{3}$ doublet ground state that carries an
electric quadrupole moment, a low lying $\Gamma_{5}$ triplet
excited state at $\sim 10$ K, and $\Gamma_{4}$ triplet and
$\Gamma_{1}$ singlet excited states at much higher energies.  This
suggests that the interaction between the quadrupole moments of
the Pr$^{3+}$ ions and the charges of the conduction electrons, as
well as the excitations between the $\Gamma_{3}$ ground state and
$\Gamma_{5}$ low lying excited state may play an important role in
generating the heavy fermion state and superconductivity in this
compound. The heavy fermion state and unconventional
superconductivity will constitute a significant challenge for
theoretical description \cite{Miyake02}.

\section{Acknowledgements}

This research was supported by US DOE Grant No.
DE-FG03-86ER-45230, US NSF Grant No. DMR-00-72125, and the NEDO
International Joint Research Program.


%
%

\end{document}